\documentclass[a4paper]{article}

\usepackage{INTERSPEECH2021}
\usepackage{amsmath,graphicx}
\usepackage{mathtools}
\usepackage{float}
\usepackage{amsfonts,amssymb}
\usepackage{booktabs}
\usepackage{subfig}
\usepackage[skip=2pt]{caption}
\usepackage[update,prepend]{epstopdf}
\usepackage[lined,algonl,ruled]{algorithm2e}
\usepackage{balance}
\usepackage{acronym}
\usepackage{color}
\usepackage{multirow}
\usepackage{multicol}
\usepackage{hyperref}
\usepackage{url}
\usepackage{pifont}

\acrodef{MoE}{Mixture of Experts}
\acrodef{SOTA}{State of The Art}
\acrodef{ATF}{Acoustic Transfer Function}
\acrodef{RIR}{Room Impulse Response}
\acrodef{SNR}{Signal-to-Noise Ratio}
\acrodef{SDR}{Signal-to-Distortion Ratio}
\acrodef{ESTOI}{Extended STOI}
\acrodef{MULCAT}{Multiply-and-Concatenate}
\acrodef{CNN}{Convolutional Neural Network}
\acrodef{SI-SNR}{Scale-Invariant Signal-to-Noise Ratio}
\acrodef{uPIT}{utterance level Permutation Invariant Training}
\acrodef{STFT}{Short-Time Fourier Transform}
\acrodef{WSJ}{Wall Street Journal}
\acrodef{RTF}{Real Time Factor}
\acrodef{MIMO}{Multiple-Inputs Multiple-Outputs}

\usepackage{bm}
\usepackage{amssymb,amsmath,graphicx,bbm,color,booktabs}
\usepackage{amsopn}

\newcommand{\ve}{\bm{e}}

\newcommand{\vech}{\bm{h}}

\newcommand{\vn}{\bm{n}}

\newcommand{\vs}{\bm{s}}       \newcommand{\vsh}{\hat{\bm{s}}}

\newcommand{\vx}{\bm{x}}

\renewcommand{\S}{\mathcal{S}}

\renewcommand{\eqref}[1]{Eq.~(\ref{#1})}

\title{Online Self-Attentive Gated RNNs \\ for Real-Time Speaker Separation}

\name{Ori Kabeli$^{1*}$, Yossi Adi$^{1*}$, Zhenyu Tang$^3$, Buye Xu$^2$, Anurag Kumar$^2$\thanks{$*$equal contribution}}
\address{
  $^1$Facebook AI Research, TLV, Israel\\
  $^2$Facebook Reality Labs, Redmond, WA, USA\\
  $^3$University of Maryland, College Park, MD, USA}
\email{\{orik, xub, anuragkr90, adiyoss\}@fb.com, zhy@umd.edu}

\begin{document}
\maketitle

\begin{abstract}
Deep neural networks have recently shown great success in the task of blind source separation, both under monaural and binaural settings. Although these methods were shown to produce high-quality separations, they were mainly applied under offline settings, in which the model has access to the full input signal while separating the signal. In this study, we convert a non-causal state-of-the-art separation model into a causal and real-time model and evaluate its performance under both online and offline settings. We compare the performance of the proposed model to several baseline methods under anechoic, noisy, and noisy-reverberant recording conditions while exploring both monaural and binaural inputs and outputs. Our findings shed light on the relative difference between causal and non-causal models when performing separation. Our stateful implementation for online separation leads to a minor drop in performance compared to the offline model;  0.8dB for monaural inputs and 0.3dB for binaural inputs while reaching a real-time factor of 0.65. Samples can be found under the following link: {\color{magenta}\url{https://kwanum.github.io/sagrnnc-stream-results/}}.
\end{abstract}

\section{Introduction}
\label{sec:intro}

In real acoustic environments, a speech source of interest is usually corrupted by interfering sounds. The human auditory system excels at attending to a target speech source, where the cocktail party problem~\cite{cherry1953some} aims to develop such capabilities in intelligent devices and systems. An important aspect of the cocktail party problem is speaker separation which aims to separate multiple concurrent speech signals of interest from a sound mixture. In \emph{blind source separation}, the condition and the scene of the mixed sources are unknown to the separation system. Recently, research on speaker source separation has seen a great leap in performance due to the success of deep learning models considering both frequency domain~\cite{hershey2016deep, yu2017permutation, chen2017deep, wang2018alternative, wang2019deep}, and time-domain~\cite{luo2019conv, stoller2018wave, venkataramani2018end, luo2020dual, zhang2020furcanext, zeghidour2020wavesplit, subakan2021attention} modeling.

\begin{figure}[t!]
\centering 
\includegraphics[width=0.8\columnwidth]{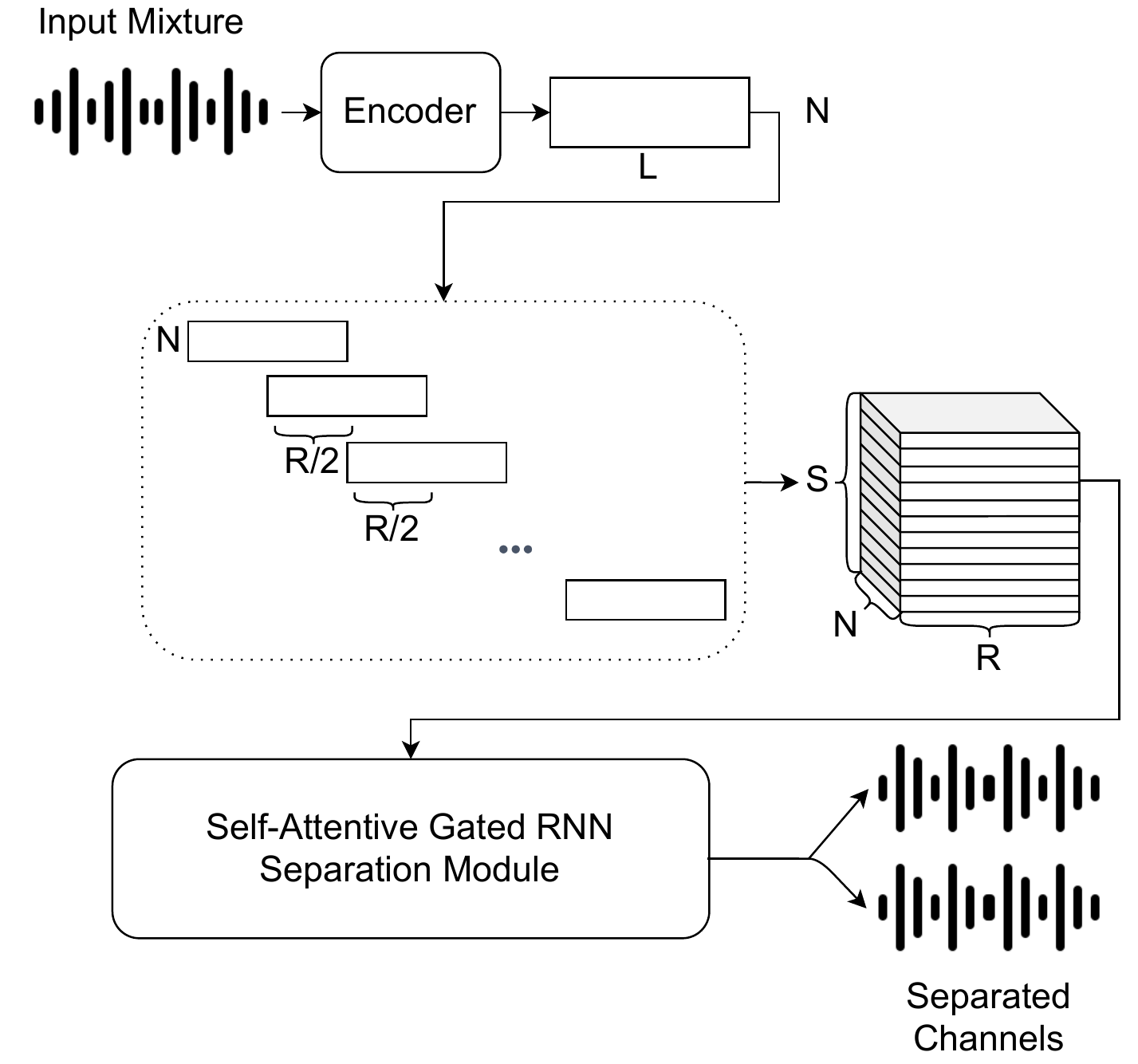}
\caption{An illustration of the separation model. Input audio is encoded into $N$ dimensional frames, which next are grouped into segments of length $R$ and chunked with an overlap of 50\%. Chunks are processed by the Self-Attentive Gated RNN module, and their output is decoded into two separate speaker streams.}
\label{fig:online}
\end{figure}

To apply source separation models to real-time systems, (e.g., VR headsets) these should have the ability to process and separate sources in an online fashion (i.e., the model separates current mixture samples without having access to future samples). However, despite the success of prior works, it mostly considers processing the input speech in an offline manner via non-causal models (i.e., the model has access to the full input speech signal before performing separation). Specifically, \ac{SOTA} models such as the one proposed in~\cite{luo2020dual, nachmani2020voice, chazan2021single, tan2020sagrnn, subakan2021attention} were developed using the inter- and intra- chunking operations together with bidirectional LSTMs. This makes the conversion to real-time and online-streaming evaluation challenging due to the dependencies in the latent representation structure. Figure~\ref{fig:online} provides a visual description of the latent representation structure. Notice that the chunking operation involves overlapping between segments, which requires a buffering and a synchronization mechanism to execute the RNNs correctly. In addition, a proper historical context of chunks needs to be managed for the self-attention module.

In this work, we study and analyze the conversion of the self-attentive gated-RNN model proposed by~\cite{tan2020sagrnn} to a real-time and streaming mode, where we consider both monaural and binaural inputs and outputs. We evaluate the discrepancy between offline, causal, and online-streaming models and under anechoic, noisy, and noisy-reverberant settings. Our results shed light on the relative difference between online and offline models when performing separation. We observe a drop of $\sim$3dB on average in terms of Signal-to-Noise Ratio improvement over the mixture. Moreover, following our stateful implementation for the online separation model, we observe a drop of less than 0.8dB for monaural inputs and less than 0.3dB for binaural inputs while reaching a real-time factor of $\sim$0.65.

\paragraph*{Related Work}
The task of online source separation was long studied under various settings~\cite{mukai2003real, togami2011online, simon2012general, von2019all, luo2018tasnet, luo2019conv, wang2019online, wu2020end, wang2020online, han2020real}. The authors of~\cite{togami2011online, simon2012general} separating the input in a sliding window while leveraging the Expectation-Maximization algorithm to estimate the model parameters. This leaves an open question of \emph{who will these separation models perform under the online setting?} To answer this question one must first design an online version of the models proposed by~\cite{luo2020dual, nachmani2020voice, chazan2021single, tan2020sagrnn}. 

The authors in~\cite{luo2018tasnet, luo2019conv, wang2019online} proposed causal and real-time separation models, however, their performance is not on par with the models proposed by~\cite{luo2020dual, chazan2021single, subakan2021attention}. In~\cite{wang2020online} the authors suggested a speaker-aware online separation method, in which the authors include an additional speaker identification loss. Although this method shows impressive results, it is orthogonal to our approach since it is based on the causal Conv-TasNet model. The authors in~\cite{von2019all} presented neural network-based models for speaker separation, counting, and diarization for meeting analysis. However, the authors evaluated their method on a single channel and anechoic setting only. Recently, the authors of~\cite{wu2020end} proposed an online separation model under the multi-channel setting, while the authors in~\cite{han2020real} suggested a modification of the TasNet model~\cite{luo2018tasnet} for binaural inputs and outputs. In section~\ref{sec:results} we empirically demonstrate that the proposed method is superior to this method under all recording conditions.

The rest of the paper is organized as follows: in section~\ref{sec:prob} we detail all the notations used throughout the paper, while in section~\ref{sec:background} we provide a background description of the evaluated models. In section~\ref{sec:method} and section~\ref{sec:results} we present the proposed method and evaluate its performance against several baseline methods. Lastly, in section~\ref{sec:con} we conclude while pointing out possible future work.

\section{Problem Setting}
\label{sec:prob}

\paragraph*{Anechoic room.}
Consider a recording mixture of $C$ different sources $\vs^j \in \mathbb{R}^T$, where $j \in \left[1,\ldots,C\right]$ in an anechoic enclosure where the source length, $T$ can vary. The mixed signal is therefore:
\begin{equation}
    \vx=\sum_{j=1}^{C} \alpha^j\cdot \vs^j,
    \label{eq:mix_anechoic}
\end{equation}
where $\alpha^j$ is the scaling factor of the $j$-th source. Although this model is commonly used to demonstrate separation abilities, anechoic noiseless environments are hard to find in the real world. 

\paragraph*{Noisy room.}
To better model real world condition, we additionally consider an additive background noise. As a results Equation~\eqref{eq:mix_anechoic} is modified to: 
\begin{equation}
    \vx=\sum_{j=1}^{C} \alpha^j\cdot \vs^j + \vn,
    \label{eq:mix_noisy}
\end{equation}
where $\vn$ is a non stationary additive noise in an unknown \ac{SNR}. Such modeling better capture real world settings, however it assumes no reverberation. 

\paragraph*{Noisy reverberant room.}
Lastly, to simulate a real-world including reverberation an \ac{ATF} which relate the sources and the microphones is considered together with additive noise as follows: 
\begin{equation}
    \vx=\sum_{j=1}^{C} \alpha^j\cdot \vs^j*\vech^j + \vn,  
    \label{eq:mix_echoic}
\end{equation}
where $\vech^j$ is the \ac{ATF} of the $j$-th source to the microphone. 

Under all three cases, we focus on the fully supervised setting, in which we are provided with a training set $\S = \{\vx_i, (\vs_{i}^1, \cdots, \vs_{i}^C)\}_{i=1}^m$, and our goal is learn a model that given an unseen mixture $\vx$, estimates $C$ separate channels, $\vsh$. In this study, we evaluate both monaural and binaural speaker separation, under the monaural setting we maximize the widely used \ac{SI-SNR}~\cite{luo2019conv} to the ground truth signals when considering reordering of the output channels $(\vsh^{\pi(1)}, \cdots, \vsh^{\pi(C)})$ for the optimal permutation $\pi$. The SI-SNR is defined as: 
\begin{equation}
    \text{SI-SNR}(\vs^j,\vsh^j)=10\log_{10}\frac{||\Tilde{\vs}^j||^2}{\|\Tilde{\ve}^j\|^2}, 
\end{equation} 
where $\Tilde{\vs}^j=\frac{\langle \vs^j,\vsh^j\rangle \vs^j}{||{\vs}^j||^2}$ and $\Tilde{\ve}^j=\vsh^j-\Tilde{\vs}^j$. 

Under the binaural setting, we use the plain SNR (also with the optimal permutation) rather than the SI-SNR as the training objective. The rationale is that SI-SNR training cannot preserve the interaural cues in the binaural estimates, as the power scale of the estimated signals is insusceptible to training due to the scale invariance. 

\section{Background}
\label{sec:background}

Recall, our goal is to evaluate recent \ac{SOTA} methods for blind source separation under the online setting. Here we briefly describe the models used in this study.

For the single-channel setting, we use a similar model to the one proposed by~\cite{nachmani2020voice, chazan2021single} equipped with a self-attention mechanism, while in the binaural setting we follow the model proposed in~\cite{tan2020sagrnn}. Both models are generally comprised of three main components: \emph{encoding and chunking}, \emph{block processing}, and \emph{decoding and overlap-add}. In the first stage, a time-domain input mixture is transformed into a set of overlapped chunks via encoding and chunking, which leads to a 3-D embedding tensor. Subsequently, the 3-D embedding tensor is passed into stacked self-attentive gated-RNN blocks to perform intra-chunk (local) and inter-chunk (global) modeling alternately and iteratively. The 3-D representation learned by the last RNN block is decoded and then transformed back to the time domain by an overlap-add operator.

\subsection{Monaural Separation}
Given a $T$-sample input waveform $\vx \in \mathbb{R}^T$, an encoder is used to segment and encode $\vx$ into $L$ overlapped time frames with a frame size of $P$ and a hop size of $P/2$, yielding a 2-D embedding $\mathbf{U} \in \mathbb{R}^{N \times L}$. Specifically, the encoder consists of a 1-D strided convolutional layer with $N$ output channels, followed by a rectified linear activation function. We divide the time frames into $S$ overlapped chunks with a chunk size of $R$ and a hop size of $R/2$. These chunks are then concatenated into a 3-D embedding $\widetilde{\mathbf{W}} = [\mathbf{W}_1, \dots, \mathbf{W}_S] \in \mathbb{R}^{N \times S \times R}$, where $\mathbf{W}_1, \dots, \mathbf{W}_S \in \mathbb{R}^{N \times R}$ are the 2-D chunks. Subsequently, the 3-D embedding $\widetilde{\mathbf{W}}$ is fed into a series of $B$ gated-RNN blocks, as proposed in~\cite{nachmani2020voice}. 

Similar to~\cite{nachmani2020voice, chazan2021single}, we use a multi-scale loss for training, which necessitates producing a waveform estimate for each speaker after each block. We decode the output embedding of each block with a decoder, which comprises a parametric rectified linear function~\cite{he2015delving} followed by a 2-D 1$\times$1 convolutional layer with $C\cdot N$ output channels. The decoded feature of size $CN \times S \times R$ is divided into $C$ 3-D representations of size $N \times S \times R$, corresponding to the $C$ speech sources. These 3-D representations are transformed back to waveforms by two successive overlap-add operations at the chunk level and the frame level, respectively.

Similarly to~\cite{tan2020sagrnn} we apply a self-attention~\cite{vaswani2017attention} mechanism before feeding the input into each gated-RNN block. We first divide a 3-D representation into a set of 2-D slices $\mathbf{Z} \in \mathbb{R}^{M \times N}$, where $M = R$ for intra-chunk modeling and $M = S$ for inter-chunk modeling. Each slice is linearly projected to a query matrix $\mathbf{Q}$, a key matrix $\mathbf{K}$ and a value matrix $\mathbf{V}$ by three different projection layers, where $\mathbf{Q}, \mathbf{K}, \mathbf{V} \in \mathbb{R}^{M \times D}$ and $D$ is set to 64. We apply a scaled dot-product attention function:

\begin{equation}
Attention(\mathbf{Q}, \mathbf{K}, \mathbf{V}) = SoftMax(\frac{\mathbf{Q}\mathbf{K}^\top}{\sqrt{D}}) \mathbf{V},
\end{equation}
where $SoftMax(\cdot)$ denotes the softmax function across columns. The output of the attention function is computed as a weighted sum of the values, where the weight assigned to each value is derived by measuring the similarities between the queries and the keys. Subsequently, all the attention output slices are merged and then linearly projected back to the size of the input 3-D representation. With a skip connection, this representation is concatenated with the input to the self-attention block and then projected back to the original size. The use of self-attention is motivated by its great success in improving separation and sequence modeling for binaural mixtures~\cite{tan2020sagrnn}. 

\subsection{Binaural Separation}
For binaural separation, we follow the same model as proposed in~\cite{tan2020sagrnn}. In which a \emph{reference encoder} and a \emph{non-reference encoder} are employed to process the binaural mixture waveforms. The resulting 2-D embeddings are concatenated and then linearly projected to the size of $N \times L$. Subsequently, we successively perform block processing, decoding, and overlap-add, akin to the monaural setting. Notice, the separation outputs always correspond to the reference ear. In order to get a multiple-inputs multiple-outputs system we alternately treating each ear as the reference. Specifically, the separation outputs for the left ear are obtained by treating the left ear as the reference ear and the right ear as the non-reference. The separation outputs for the right ear are obtained by swapping the inputs of the two ears. Note that the same system is used for separation in both channels. Such a cross-ear referencing strategy selects the target channel by exploiting discriminative information within the ordered pair of channels.

\section{Method}
\label{sec:method}

Recall, our model is comprised of a 1-D convolutional encoder, a chunking operation, which converts the 2D input into a 3D tensor of overlapping segments, and a series of self-attentive gated RNNs.  Originally, these models were designed to work in a non-causal manner, i.e., see the full input sequence in advance. 

Converting the model into its causal version can be straightforward: i) converting the encoder into a causal convolution, ii) changing all RNNs into uni-directional and iii) replacing the self-attention with a causal self-attention, similarly to the one proposed in~\cite{merity2019single}. However, it remains unclear how to modify the chunking operation under real-time and online settings. Since the RNN blocks are being applied over both chuncking dimensions (i.e., $R$ and $S$) which consists of a 50\% overlap, the RNN which processes the $R$ dimension sees future context. Hence a synchronization scheme is needed to fully convert it into an online and real-time separation model. In the next subsections, we suggest both a \emph{Stateless} and a \emph{Stateful} approaches for operating the model in a real-time and streaming manner. 

\begin{figure}[t!]
\centering 
\includegraphics[width=0.8\columnwidth]{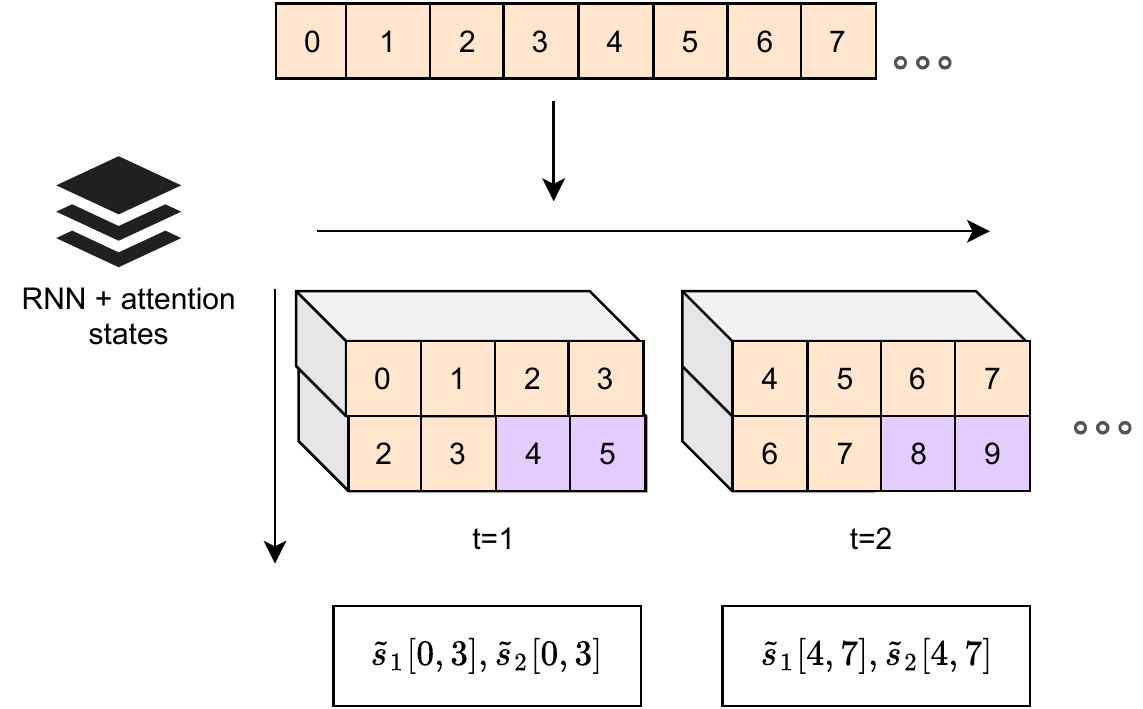}
\caption{An illustration of the stateful streaming mode, shown for simplicity with a segment size $R=4$. Input audio is buffered until it reaches the length of $1\frac{1}{2} \times R$. Then a segment and an additional half-segment future context (squares [0-5] in illustration) are pulled from the buffer. These $1\frac{1}{2} \times$ segments are chunked and processed by the SAGRNN module. Processed chunks are merged together into an output segment (squares [0-3] in illustration).}
\label{fig:stateful} 
\end{figure}

\subsection{Stateless Mode}
Our first evaluation uses a naive implementation of audio streaming in which the model does not store any state between processing audio segments, denoted as \emph{stateless}. Under the stateless approach, the model is being fed with audio chunks, each of which contains a segment of audio, future context, and historical context. The historical context is crucial for better modeling long-range dependencies to improve the separation results. 

In this mode, the model forward function is idempotent. Thus, no state is retained within neither the RNNs nor the processing of the segment (chunking, merging, etc.). The audio chunks are processed using a sliding window with a hop size of $R$. The future context equals $R/2$ to accommodate for chunks overlap, and we set the historical context to be 640ms. We experimented with several lengths of historical context and found 640ms to perform the best while maintaining a sensible RTF of close to real-time.

Despite its simplicity and ease of implementation, this approach wastes computation and produces channel estimation with significantly lower quality. Moreover, our results show that even with a large historical context of ($>$640ms), in which the \ac{RTF} is greater than 1 ($\sim$ 1.3), the results are still not on par with the causal, non-streaming model.

\subsection{Stateful Mode}
Next, we design the stateful-model-based approach. With it, audio segments are fed into the model in a streaming manner. The model maintains an internal state between each forward calls to keep track of historical context. We maintain a separate state for each of the RNNs and self-attention blocks. Notice, our GatedRNN block is composed of two separable RNN blocks operating on different dimensions. At inference time, the $R$ dimension stays fixed while the $S$ dimension varies as a function of the input length ($S \approx \lfloor 2L / R\rfloor - 1$). Hence, we need to maintain a state only for the RNNs which processing inputs over the $S$ axis.

Recall, we reconstruct back the signal using an overlap-and-add operation. Due to the 50\% overlap between chunks, our method needs to store the future context of half a segment (i.e., $R/2$). As a result, to process segment $i$ the first half of segment $i+1$ is buffered as the future context. Formally, to estimate the separated sources for segment $i$ our model gets as input the $i_{th}$ segment of size $R$, together with a future context of half a segment of size $R/2$, denoted by $\textit{seg}_{i}$. The RNNs process $\textit{seg}_{i}$ and output $\tilde{s}_{i}$, which is followed by an overlap-and-add operation to estimate the waveform. 

Next, to further process the $i+1_{th}$ segment accurately, our method keeps states for the RNN hidden states, and historical context for the self-attention. Notice, the future context from segment $i$ is the first half of the segment $i+1$. A visual example can be found in Figure~\ref{fig:stateful}. For simplicity, we demonstrate the model’s operation with segment size $R=4$. We buffer the input audio until it reaches the length of $1\frac{1}{2} \times R$. Then, we process the first segment and keep the future context in the buffer for further processing. 

Using the stateful method, we demonstrate that results meet the real-time requirement of an \ac{RTF} lower than 1 ($\sim$0.65) while preserving the audio quality almost intact, as tested under several settings and objective measurements.

\subsection{Model latency}
The latency of the streaming models depends on three terms: i) \ac{RTF}; ii) processed segment size; iii) and future context. Formally, we measure the latency as follows,  
\begin{equation}
Latency = R + RTF * R + FC, 
\label{eq:latency}
\end{equation}
where $FC$ is the future context. Our chosen segment size corresponds to 64ms of audio at a sample rate of 8Khz, where on each model forward pass the model processes one segment of audio. The internal buffers of the model keep a future context that is defined to be 32ms (to account for future half-segments needed to process the current segment). Overall, the latency of the proposed method is on average 138ms, when considering \ac{RTF} of 0.65 (64 + 64*0.65 + 32).

\section{Results}
\label{sec:results}

We evaluated the proposed streaming implementation considering both monaural and binaural settings. For the monaural experiments, we used the wsj2-mix~\cite{hershey2016deep}, WHAM!~\cite{wichern2019wham}, and WHAMR!~\cite{maciejewski2020whamr} as the anechoic, noisy, and noisy reverberant settings respectively. 

Under the binaural setup, we used the noise-free and noisy datasets as suggested in~\cite{tan2020sagrnn}. To create the noisy-reverberant dataset, we generally followed the procedure of the noisy datasets, except that the simulated binaural room impulse responses (BRIRs), instead of the head-related impulse responses (HRIRs), were used to be convolved with the speech and noise signals to simulate room acoustics effects. The BRIRs were created in two steps: 1) room acoustics simulation and 2) HRIR convolution. Given a room and a source location, we simulated the sound field at the listener's position based on a combination of image-source and ray-tracing approaches~\cite{rindel2000odeon, scheibler2018pra}. The image-source method (up to six orders of reflections) was used to simulate the sound pressure of the direct sound and the early reflections at the center location of the listener's head. Each incident sound wave was then convolved with the listener's HRIRs corresponding to the incident angle to compose the early portion of the binaural impulse response. The ray-tracing method was used to simulate the energy flow of the late reverberation (beyond six orders of reflections). The direction, timing, and energy information of the acoustic energy flow were used to create the late portion of the BRIR based on the head orientation and the averaged distance between the left and right ears. Ten thousand ``shoebox" rooms were created for simulation, with each length, width, and height being randomly sampled between 2m and 10m. Frequency-independent absorption coefficients were randomly assigned to the walls. And we select a set of impulse responses that have the reverberation time uniformly distributed between 0.1s and 1s. In each room, the source and listener locations were also randomly sampled. Different HRIRs from the CIPIC HRTF Database~\cite{algazi2001cipic} were randomly selected to generate BRIRs for each room. 

Under all settings, we consider two speakers only in the input mixtures. Each dataset contains 20,000, 5,000, and 3,000 mixtures in the training, validation, and testing sets, respectively.

\subsection{Monaural Speaker Separation}
We start by evaluating the proposed method for single channel separation. We compared our method against LSTM-TasNet~\cite{luo2018tasnet} (causal), Conv-TasNet~\cite{luo2019conv} (both causal and non-causal), Dual-Path RNN (DPRNN)~\cite{luo2020dual} (non-causal), and GatedRNN~\cite{nachmani2020voice} (both causal and non-causal). Table~\ref{tab:single_channel} summarizes the results. 

Converting the SAGRNN model to be causal (denoted as SAGRNNc) yields a drop of 3.6dB, 2.6dB, and 2dB on the anechoic, noisy, and noisy-reverberant settings respectively in terms of \ac{SI-SNR} improvement over the mixture. These results are superior by a significant margin to the causal baseline methods. Then, we evaluate the SAGRNN model under the online setting when considering both stateless and stateful modes. Results are presented on Table~\ref{tab:single_channel} (bottom).

\begin{table}[t!]
	\caption{Results for single channel speaker separation using causal and non-causal models under various recording conditions. Results are reported in terms of SI-SNR improvement over the mixture in dB. We additionally compare results from non-streaming, stateful and stateless streaming modes.}
	\centering
	\resizebox{0.48\textwidth}{!}{
		\begin{tabular}{|l|c|c|c|c|c|}
			\hline
			 & \# param. & Anechoic & Noisy & Noisy reverberant & Causal \\
			\hline
			Conv-TasNet~\cite{luo2019conv}          & 5.1~M  & 15.3  & 12.7   & 8.3    & \ding{55} \\ 
			DPRNN~\cite{luo2020dual}                & 3.6~M  & 18.8  & 13.9   & 10.3   & \ding{55} \\
			SAGRNN                                  & 7.6~M & 19.7   & 15.2   & 12.2   & \ding{55} \\
			\hline
			TasNet~\cite{luo2018tasnet}             & 32.0~M & 10.8  & 8.7    & 7.2    & \ding{51} \\ 
			Conv-TasNet~\cite{luo2019conv}          & 5.1~M  & 10.6  & 8.6    & 7.2    & \ding{51} \\
			SAGRNNc                                 & 4.7~M  & 16.1  & 12.6   & 10.2   & \ding{51} \\
			\hline
			SAGRNNc (\emph{Stateless})            & 4.7~M  & 9.5   & 7.3     & 7.0   & \ding{51}\\
			SAGRNNc (\emph{Stateful})             & 4.7~M  & 15.3  & 12.4    & 10.1  & \ding{51}\\
			\hline
		\end{tabular}}
	\label{tab:single_channel}
\end{table}

\begin{table*}[t!]
	\caption{Binaural speaker separation results using causal and non-causal models under various recording conditions. Results for SDRi and SNRi are reported in dB while ESTOI results are reported in percentage.}
	\centering
	\resizebox{\textwidth}{!}{
		\begin{tabular}{|l|c|c|c|c|c|c|c|c|c|c|c|c|c|c|}
			\hline
			 & SDRi & SNRi & ESTOI & PESQ & SDRi & SNRi & ESTOI & PESQ & SDRi & SNRi & ESTOI & PESQ & Causal\\
			\hline
			& \multicolumn{4}{c|}{Anechoic} & \multicolumn{4}{c|}{Noisy} & \multicolumn{4}{c|}{Noisy reverberant} & \\
			\hline
			MIMO TasNet~\cite{han2020real}  & 21.14 & 20.69 & 95.53 & 3.73 & 14.40 & 15.23 & 63.79 & 2.41 & 3.95 & 7.62 & 27.81 & 1.73  & \ding{51}\\
			Oracle MB-MVDR                  & 17.13 & 10.44 & 95.77 & 3.66 & 4.98 & 4.90 & 42.71 & 1.79 & 1.74 & 4.01 & 29.44 & \ 1.79  & \ding{51}\\
			MIMO SAGRNN~\cite{tan2020sagrnn} & 27.19 & 26.88 & 98.08 & 4.06 & 17.53 & 17.95 & 75.14 & 2.78 & 8.31 & 9.56 & 36.42 & 2.00 & \ding{55}\\		 
			\hline
			MIMO SAGRNNc                     & 25.73 & 24.12 & 97.2 & 3.86 & 15.04 & 15.91 & 67.11  & 2.41 & 6.37 & 8.70 & 30.51 & 1.85 & \ding{51} \\
			MIMO SAGRNNc (Stateless)         & 23.57 & 22.06 & 96.0 & 3.67 & 12.81 & 13.22 & 62.50  & 2.11 & 3.78 & 7.47 & 27.12 & 1.68 & \ding{51} \\ 
			MIMO SAGRNNc (Stateful)          & 25.63 & 24.11 & 97.0 & 3.85 & 14.73 & 15.77 & 66.52  & 2.35 & 5.41 & 8.20 & 29.55 & 1.79 & \ding{51} \\
			\hline
		\end{tabular}}
	\label{tab:bi_sep}
\end{table*}

Notice, we observe a large drop in performance under the stateless mode for all settings (e.g., 6.6dB decrease in performance for anechoic samples). These results suggest that the separation model does benefit from a large historical context. Moreover, the \ac{RTF} of the stateless approach is on average $\sim1.3$, hence does not meet the real-time requirement. However, when considering the stateful approach, the gap between the streaming and non-streaming modes is significantly smaller (e.g., 0.8 dB decrease in performance for the anechoic samples) while its \ac{RTF} is on average $\sim$0.65. 

The discrepancy in performance between the offline and stateful evaluations is caused due to: (1) the self-attention layers look at a finite buffer of historical segments to process their dependencies with the current segment of audio (as opposed to looking at all past segments). (2) The overlap-and-add operation outputs a reminder of audio that is trimmed to fit the original input audio length. The slight dislocation caused by the output of the overlap-and-add is negligible in an offline setting but becomes more apparent when iteratively processing small chunks of data.

\subsection{Binaural Speaker Separation}
Next, we evaluated the causal SAGRNN model under the binaural setting. In the following, the input is a binaural signal and the output is a binaural estimate of the separated sources. We refer to this system as \ac{MIMO}. Results are summarizes in Table~\ref{tab:bi_sep}. We report \ac{SNR} and \ac{SDR} improvement over the mixture, together with \ac{ESTOI} and PESQ. Results suggest that the causal SAGRNN is superior to the baseline methods under all settings and evaluation metrics. However, the gap between causal and non-causal models is smaller in the binaural setting than the gap in the single-channel setting. This may happen due to the multiple inputs, which provide additional tracking information to the model. 

When considering an online evaluation, we observe a similar trend in which the stateless approach performs worse than the stateful approach, however, the gap is smaller than the one under the single-channel setting. Notice, under the stateful setting, the gap between streaming and non-streaming modes is negligible. 

Interestingly, the binaural and monaural implementations show similar \ac{RTF} rations - this is due to the model implementation utilizes GPU parallelism for the processing of the different audio channels, while the processed segment size stays the same in both settings.

\paragraph*{The effect of segment size.} Lastly, we analyze the effect of the segment size, $R$. Recall, as stated in equation~\eqref{eq:latency}, the latency is a function of the segment size. Hence, in order to better understand the effect of the segment size of the model performance and latency we trained several models where $R \in \{32, 64, 128, 256\}$ corresponding to latency of $[65.6, 85.9, 142.1, 323.1]$ milliseconds respectively. For this set of experiments, we trained a binaural separation model using the anechoic datasets and report \ac{SNR} and \ac{SDR} improvement over the mixture, \ac{ESTOI}, PESQ, and \ac{RTF}. 

\begin{figure}[t!]
  \subfloat[SDRi and SNRi]{
	\begin{minipage}[c][1\width]{
	   0.23\textwidth}
	   \centering
	   \includegraphics[width=1\textwidth]{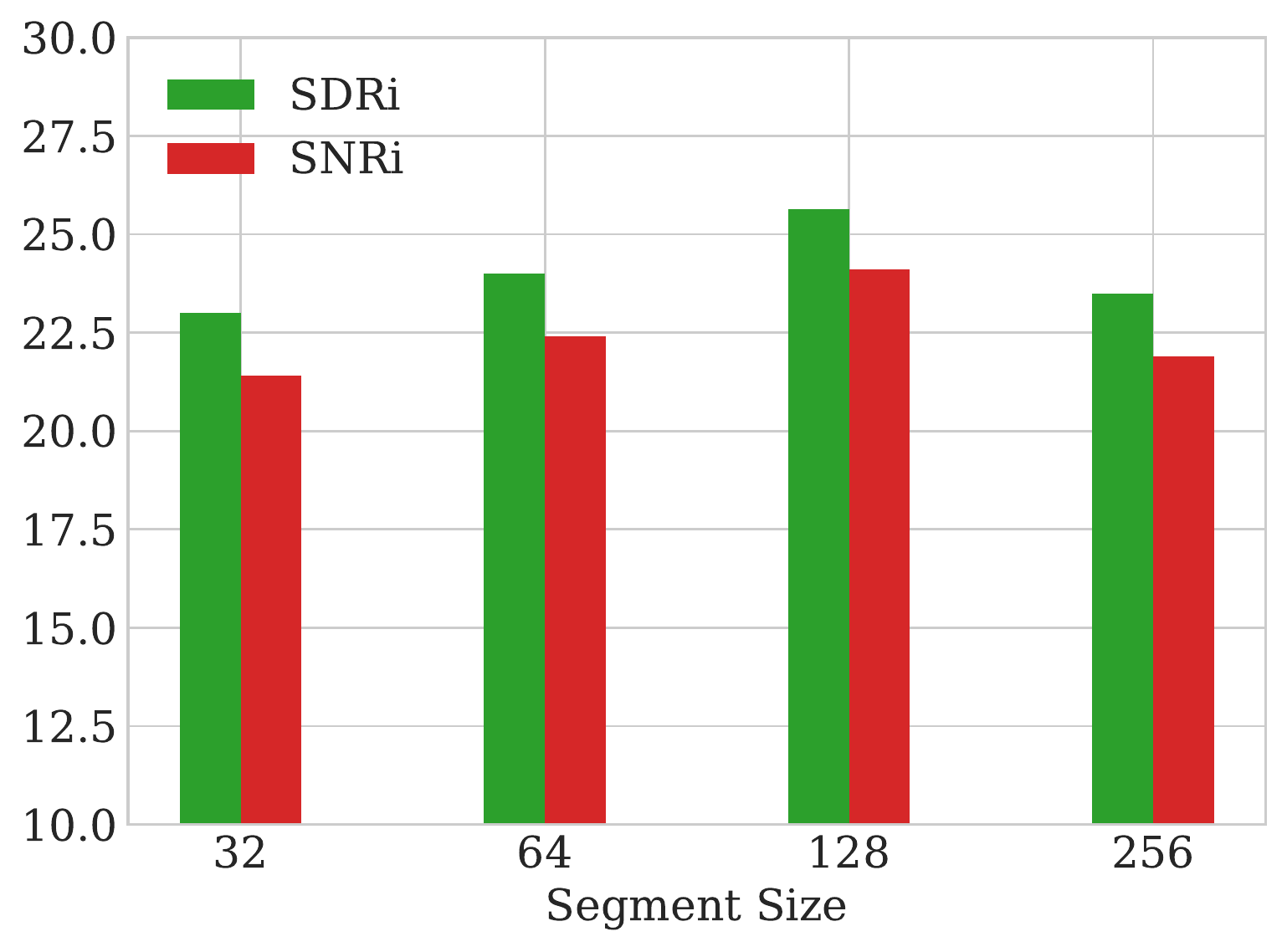}
	\end{minipage}}
  \hfill 	
  \subfloat[STOI]{
	\begin{minipage}[c][1\width]{
	   0.23\textwidth}
	   \centering
	   \includegraphics[width=1\textwidth]{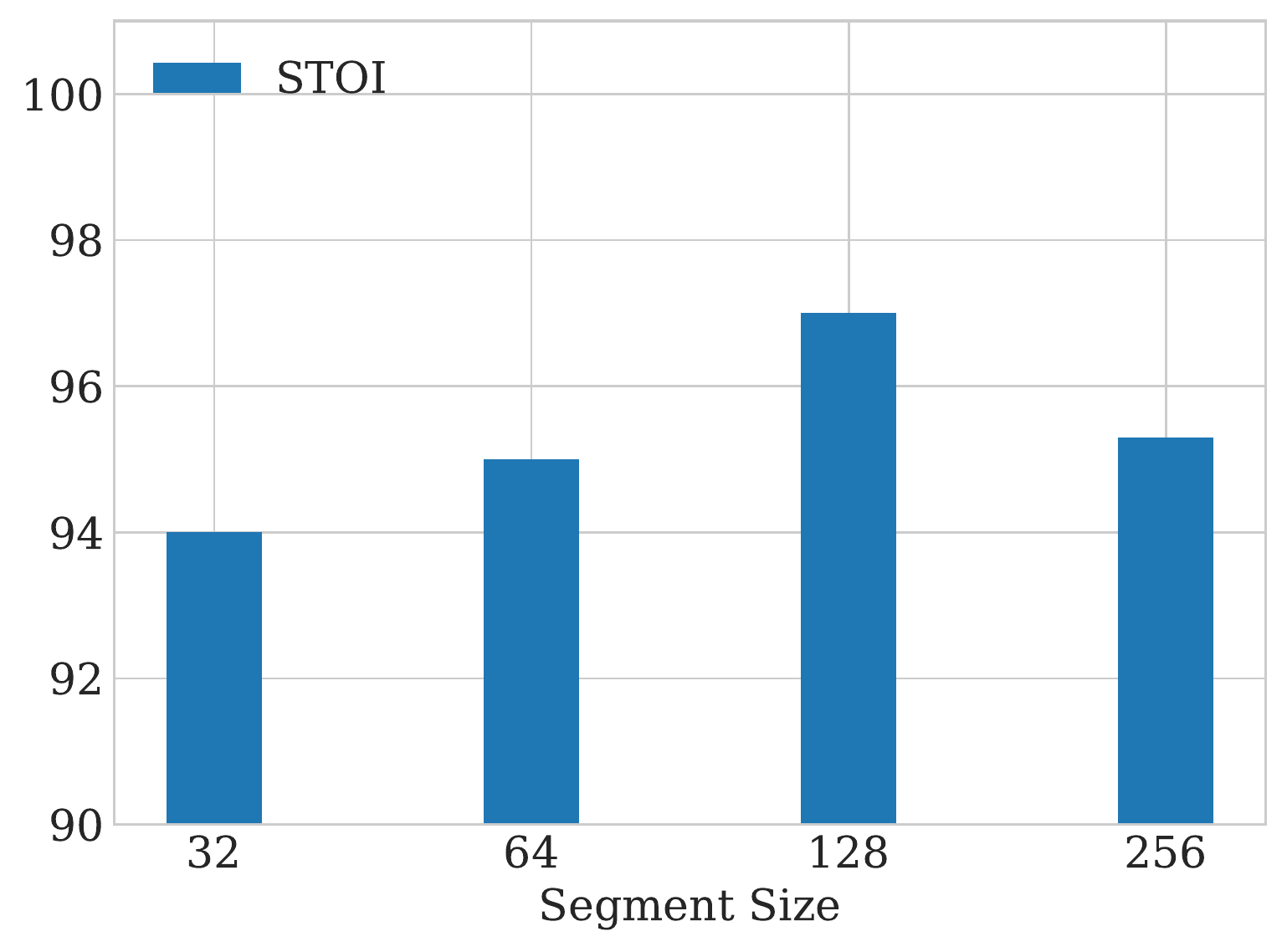}
	\end{minipage}}
 \\
  \subfloat[PESQ]{
	\begin{minipage}[c][1\width]{
	   0.23\textwidth}
	   \centering
	   \includegraphics[width=1\textwidth]{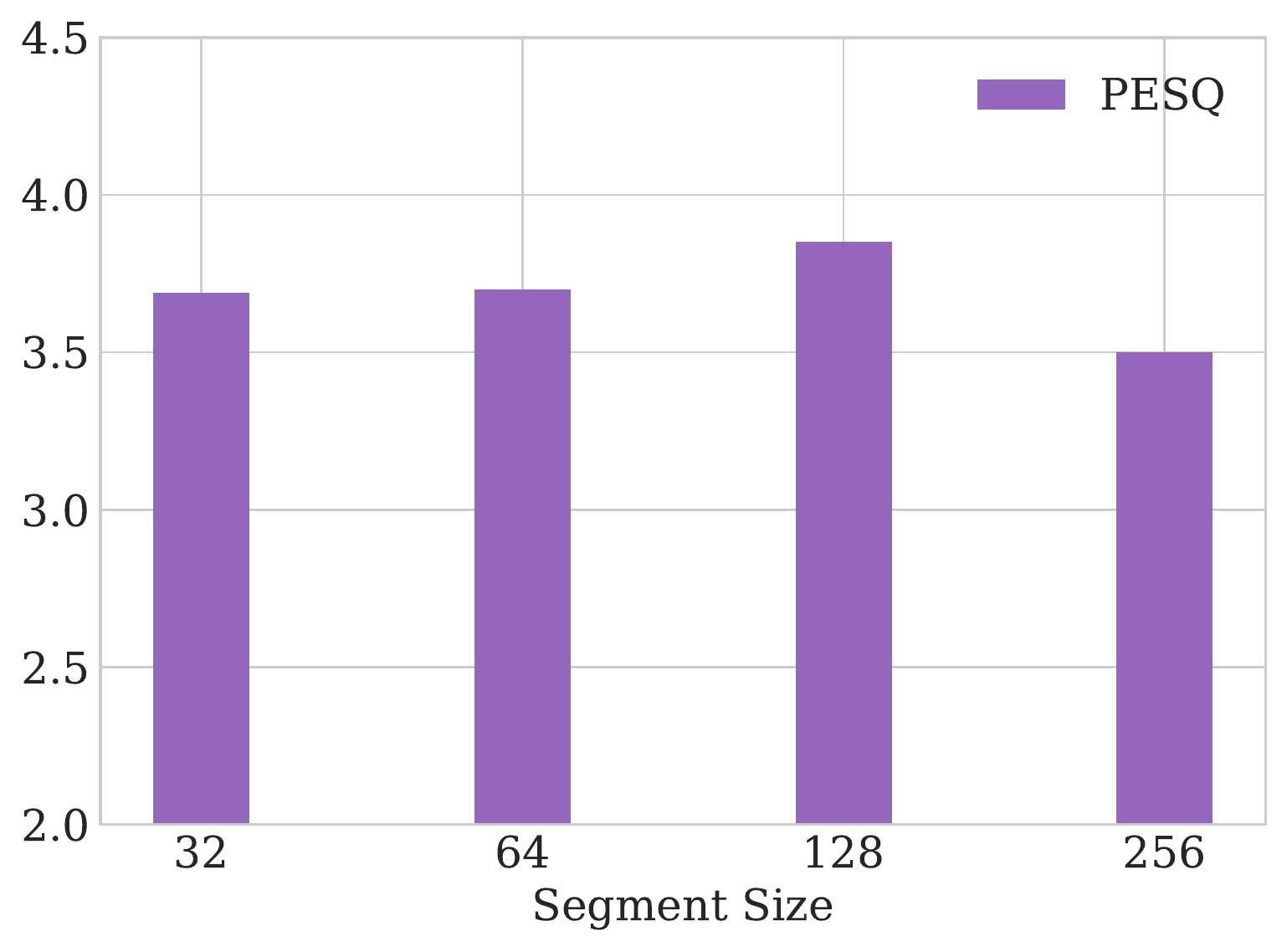}
	\end{minipage}}
 \hfill	
  \subfloat[RTF]{
	\begin{minipage}[c][1\width]{
	   0.23\textwidth}
	   \centering
	   \includegraphics[width=1\textwidth]{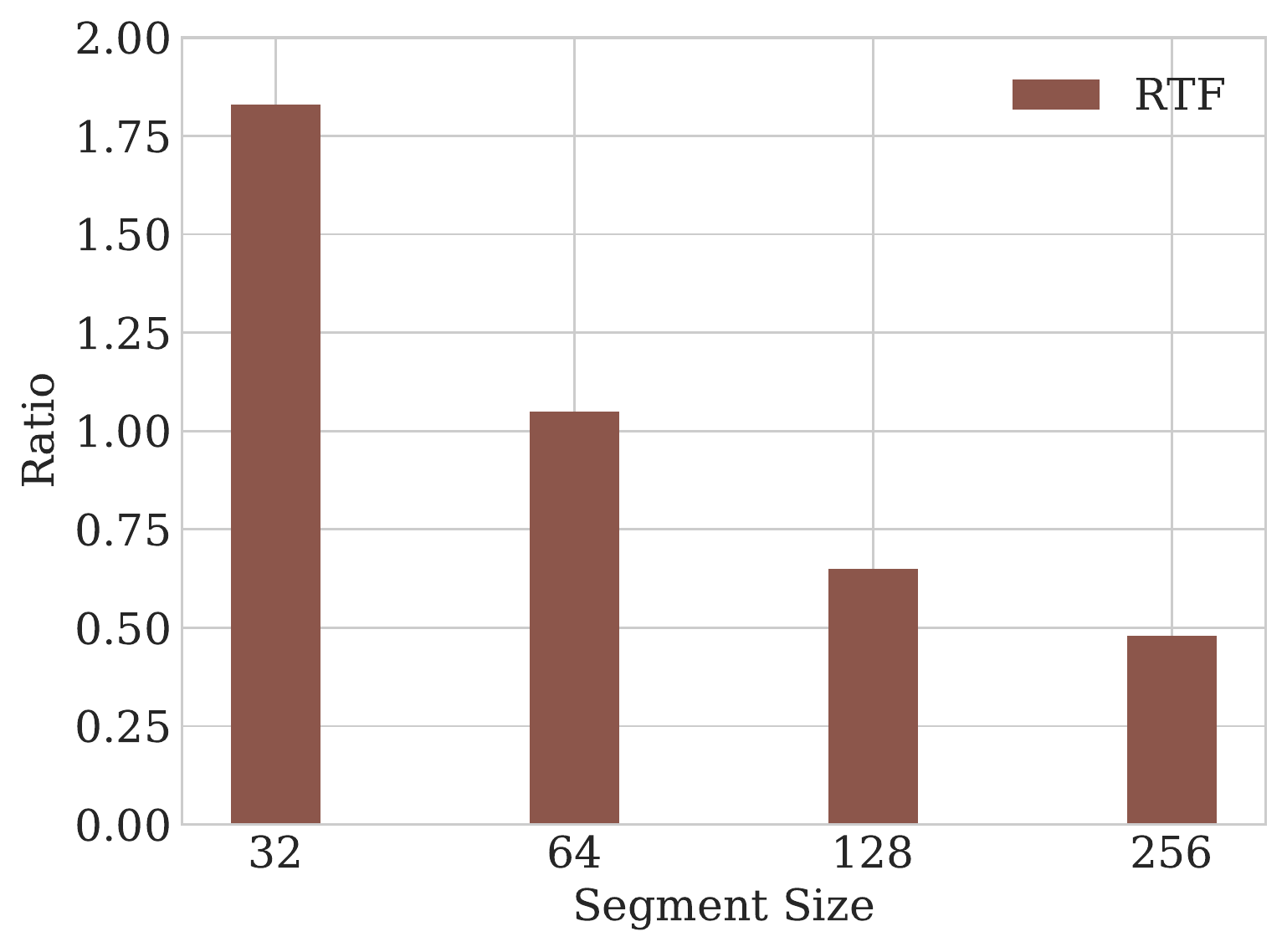}
	\end{minipage}}
\caption{Results for different segment size values ($R \in \{32, 64, 128, 256\}$). Results are reported for the clean anechoic binaural speaker separation using the stateful streaming mode. We report SNR and SDR improment over the mixture, PESQ, STOI and \ac{RTF}.}
\label{fig:seg_size}
\end{figure}

Results suggests that while setting $R=128$ we reach the best performance in terms of \ac{SDR}i, \ac{SNR}i STOI, and PESQ. When considering \ac{RTF}, $R=256$ gets better real-time ratios, however this comes at the expense of separation performance. Notice, $R=\{32, 64\}$ does not meet the real time requirements and reaching \ac{RTF} greater than 1. Figure~\ref{fig:seg_size} summarizes the results.

\vspace{-0.2cm}
\section{Conclusion \& Future Work}
\vspace{-0.1cm}
\label{sec:con}
In this work, we studied the SAGRNN model under online and real-time settings. We report results for both monaural and binaural inputs under anechoic, noisy, and noisy-reverberant recording conditions, in which we explored both stateless and stateful modes. Our findings suggest that converting the SAGRNN to a causal model costs a drop of $\sim$3dB on average for a single channel and $0.8$dB on average for the binaural setup. Moreover, when evaluating the models under the online setting, our empirical study suggests the stateful mode is superior to the stateless approach while reaching \ac{RTF} of ~0.65 on average. 

Recall, our model has a latency of 138ms, this can be further reduced using a shorter future context. For future work, we would like to explore shortening the future context and processing time of an audio segment by exploring various segments and overlapping sizes. Additionally, improving the synchronization mechanisms between the processing done by the RNNs may decrease the overall model latency. Lastly, to support commodity hardware, future work will also include an online real-time CPU implementation of the model, where trade-offs between speed and quality will be analyzed.

\clearpage
\bibliographystyle{IEEEtran}
\bibliography{refs}

\end{document}